\documentclass[a4paper]{article}

\usepackage{INTERSPEECH2021}
\usepackage{amsmath,graphicx,mathtools,amssymb,comment}
\usepackage{color}
\usepackage{bbold}
\usepackage{multirow}
\usepackage{float}
\usepackage{caption}
\usepackage{subcaption}
\usepackage{hyperref}
\usepackage{hhline, pifont, bbm}
\newcommand{\cmark}{\ding{51}}%
\newcommand{\xmark}{\ding{55}}%

\def\L{\mathcal{L}}
\def\W{\texttt{W}}
\def\U{\texttt{U}}
\def\WU{\texttt{WU}}
\def\WD{\texttt{WD}}
\def\WUD{\texttt{WUD}}
\def\uttwcr{\hat{\mu}_{\text{wcr}}}
\def\uttprob{\hat{\mu}_{\text{utt}}}
\def\uttwer{\hat{\mu}_{\text{wer}}}

\title{Multi-Task Learning for End-to-End ASR Word and Utterance Confidence with Deletion Prediction}
\name{David Qiu$^1$, Yanzhang He$^1$, Qiujia Li$^{2*}$, Yu Zhang$^1$, Liangliang Cao$^1$, Ian McGraw$^1$
\thanks{$^*$Work was done while the author interned at Google.}}
\address{
  $^1$Google, LLC, USA, $^2$University of Cambridge, UK}
\email{$^1$\{qdavid, yanzhanghe, ngyuzh, llcao, imcgraw\}@google.com, $^2$ql264@cam.ac.uk}

\begin{document}

\maketitle
\begin{abstract}
Confidence scores are very useful for downstream applications of automatic speech recognition (ASR) systems. Recent works have proposed using neural networks to learn word or utterance confidence scores for end-to-end ASR. In those studies, word confidence by itself does not model deletions, and utterance confidence does not take advantage of word-level training signals. This paper proposes to jointly learn word confidence, word deletion, and utterance confidence. Empirical results show that multi-task learning with all three objectives improves confidence metrics (NCE, AUC, RMSE) without the need for increasing the model size of the confidence estimation module. Using the utterance-level confidence for rescoring also decreases the word error rates on Google's Voice Search and Long-tail Maps datasets by 3-5\% relative, without needing a dedicated neural rescorer.
\end{abstract}

\section{Introduction}
Confidence scores are often used to assess the reliability of speech recognizers~\cite{wessel2001confidence,jiang2005confidence}. They have been widely used for various downstream tasks, including semi-supervised and active learning~\cite{wang2007unsupervised,yu2010unsupervised,yu2010active,huang2013semi,park2020improved,li2006confidence}, system combination~\cite{fiscus1997post,Evermann2000PosteriorPD,evermann2000large}, dialog systems~\cite{hakkani2004unsupervised,riccardi2005active,tur2005combining} and keyword spotting~\cite{benayed2003confidence,keshet2009discriminative,seigel2013confidence}. For traditional hidden Markov model (HMM)-based automatic speech recognition (ASR) systems, the simplest approach is to use the posterior probability for each hypothesized word. More advanced methods have been proposed to better estimate word-level confidence, including linear models~\cite{evermann2000large,gillick1997probabilistic}, conditional random fields~\cite{seigel2011combining}, and neural networks~\cite{kalgaonkar2015estimating,del2018speaker,ragni2018confidence,li2019bi}.

However, for end-to-end (E2E) ASR models such as recurrent neural network transducers (RNN-T) and attention-based sequence-to-sequence models, word posteriors cannot be approximated well from the tree-like ``lattice'' where the prediction of each token conditions on the full history of previous tokens. Autoregressive decoders also tend to be overconfident~\cite{li2020confidence}.
To solve this challenge, several model-based methods have been proposed to estimate
word and utterance-level confidence for E2E models. For examples, \cite{woodward2020confidence,li2020confidence} proposed to train a token-level (e.g. graphemes or word-pieces~\cite{Schuster2012wordpiece}) confidence estimation module (CEM) on top of a given E2E model and the word-level confidence can be simply obtained by averaging the token-level scores. Since word-level confidence is more commonly used for applications such as keyword spotting~\cite{benayed2003confidence,keshet2009discriminative,seigel2013confidence} and systems combination~\cite{fiscus1997post,Evermann2000PosteriorPD}, \cite{qiu2021learning} proposed to train CEMs for word-level confidence directly, where issues such as multiple possible word-piece tokenizations are addressed. Utterance-level confidence scores are also important for applications such as data selection~\cite{yu2010unsupervised,huang2013semi,park2020improved}. One can directly average the word-level confidence scores for utterance-level confidence, but averaging is sub-optimal since deletion errors are not accounted for. \cite{kumar2020utterance,li2021residual} proposed to directly learn utterance-level confidence scores for E2E systems and \cite{li2021residual} showed that rescoring n-best hypotheses based on utterance-level confidence estimator can also reduce word error rates (WER). Previous works generally model confidence scores at a single level (\emph{i.e.} token, word or utterance). This paper explores joint optimization of word-level and utterance-level confidence estimation in a single model with multi-task learning.

Model-based approaches typically formulate the confidence estimation problem as a binary classification task~\cite{kalgaonkar2015estimating,del2018speaker,ragni2018confidence,li2019bi,woodward2020confidence,li2020confidence,qiu2021learning,kumar2020utterance,li2021residual}, where correct tokens, words, or utterances should have confidence scores close to 1, and 0 otherwise. For word-level confidence estimation, scores between 1 and 0 are only assigned to words that appear in the hypotheses. This covers substitution and insertion errors for ASR. However, deletion errors are either ignored or not explicitly addressed. In other words, averaging the confidence scores yields an estimate for the word correct ratio (WCR), but deletion errors need to be considered for estimating WER. As shown in~\cite{Seigel2014DetectingDI,ragni2018confidence}, deletion estimation is also important for downstream tasks. For example, even with a good confidence estimator, utterances with high confidence may have high WER because of deletion errors. Tasks such as semi-supervised training can be adversely affected without explicit modeling of deletion errors~\cite{ragni2018confidence}. In this work, we explore estimating deletion errors along with word-level and utterance-level confidence scores in a single model.

This paper makes the following contributions: 1) Analyzes the word-level CEM in~\cite{qiu2021learning} to show that it mitigates the overconfidence problem when subword deletions are present. 2) Proposes using neural networks and Poisson regression to learn arbitrary deletion length, and integrate it with the word-level CEM to estimate WER. 3) Demonstrates that jointly learning the word confidence, utterance confidence, and deletion length prediction tasks improves all reported confidence metrics over single-task word-level or utterance-level CEMs. 4) Shows that using the utterance-level confidence to rescore RNN-T hypotheses reduces WERs on industry-scale datasets.

\section{Model}
\label{sec:model}
\begin{figure*}[t]
   \centering
     \includegraphics[width=0.9\linewidth]{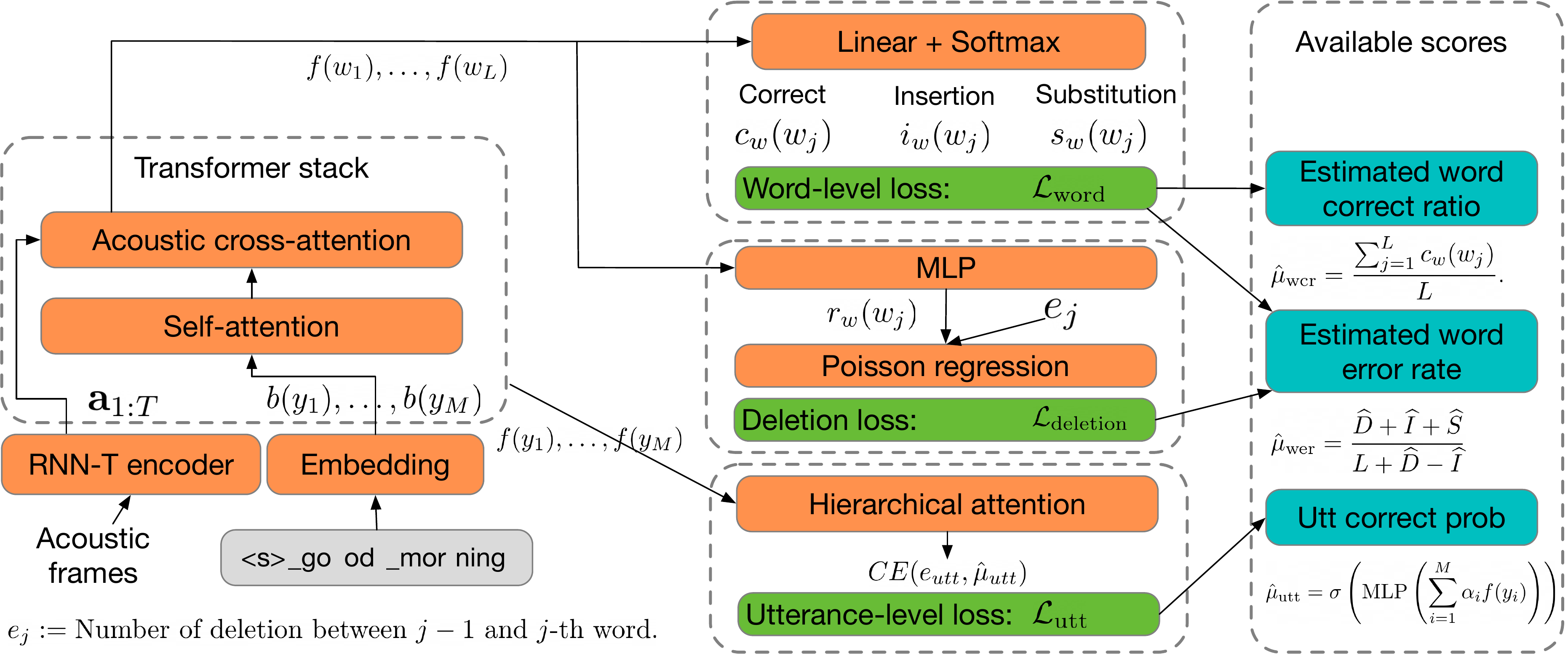}
     \caption{Model architecture. $f(\cdot)$ denotes token/word level features. $c_{w}, i_{w}, s_{w}$ denote the estimated correction, insertion and substitution. $\L_{\text{deletion}}$ refers to Poisson regression for estimating deletion errors. More details in Section \ref{sec:model}.}
     \label{Fig: Architecture}
     \vspace{-1em}
\end{figure*}

\subsection{Notation}
In general, an E2E ASR model processes a sequence of acoustic features $\mathbf{a}$, and outputs a token sequence $[y_1, \ldots, y_M]$ as its hypothesis. Popular choices for the set of tokens include graphemes and word-pieces (WP). In this paper, we focus on WP. The goal of the confidence model is to use both $\mathbf{a}$, and the embedding of the WP sequence, $\mathbf{b}$, as inputs, and output a score for every word and every utterance. To distinguish between WP and words, we use $M$ and $L$ to refer to the number of WPs and words in a hypothesis, respectively. We use $y_i$ and $w_j$ to denote the $i$-th WP and $j$-th word, respectively.

We use the Transformer~\cite{vaswani2017attention, li2020parallel} as the base feature extractor for all of the proposed tasks:
\begin{align}
    [f(y_1), \ldots, f(y_M)] &= \mathrm{Transformer}(\mathrm{CA}(\mathbf{a}), \mathrm{SA}(\mathbf{b})), \label{Eq: Feature}
\end{align}
where the cross-attention block $\mathrm{CA}$ attends to $\mathbf{a}$, and the self-attention block $\mathrm{SA}$ attends to $\mathbf{b}$. We refer to the Transformer's output at the $i$-th token as $f(y_i)$.

\subsection{Baseline Word-Level Confidence Estimation}

For the baseline CEM, we use a multi-layer perception (MLP) to process the Transformer's output $f$ into 3-D outputs:
\begin{align}
    [c(y_i), i(y_i), s(y_i)] &= \mathrm{Softmax}(\mathrm{MLP}(f(y_i))). \label{Eq: Softmax}
\end{align}
For training, we adapt the word-level training loss introduced in~\cite{qiu2021learning}. To recap, the loss gathers the output of~\eqref{Eq: Softmax}, but only at the indices corresponding to the last WP of each word (see Table~\ref{Table: Edit}). The gathered output can be re-indexed as $[c_w(w_1), \ldots, c_w(w_L)]$, $[i_w(w_1), \ldots, i_w(w_L)]$, and $[s_w(w_1), \ldots, s_w(w_L)]$, to denote the probabilities that the words are tagged by Levenshtein edit distance~\cite{levenshtein1966binary} as ``correct'', ``insertion'', and ``substitution'', respectively. Then, over every word, we compute the cross-entropy loss $\L_\text{word}$ between the ground truth tags and the outputs:
\begin{align}
    \label{Eq: Word loss}
    -\sum_{j=1}^L \left[ \mathbbm{1}^\text{cor}_j \log c_w(w_j) + \mathbbm{1}^\text{ins}_j \log i_w(w_j) + \mathbbm{1}^\text{sub}_j \log s_w(w_j) \right].
\end{align}
This also yields a simple estimate for the WCR: 
\begin{align}
    \label{Eq: WCR}
    \uttwcr = \frac{\sum_{j=1}^L c_w(w_j)}{L}.
\end{align}

\begin{table}[ht]
\caption{Top: example of the WP edit distance outputting ``correct'' for a WP when the whole word is actually a ``substitution'' error. Bottom: example of computing the word-level loss at the last WP of each word.}
\vspace{-1em}
\label{Table: Edit}
\centering
\begin{tabular}{ l@{\hspace{1\tabcolsep}}cccc}
 \toprule
 Hyp: & \_go &  & \_morn & ing \\ 
 Ref: & \_go & od & \_morn & ing \\ 
 WP edit: & \textit{cor} & \textit{del} & \textit{cor} & \textit{cor} \\ 
 Word edit: & \textit{sub} & -- & -- & \textit{cor} \\
 \midrule
    $\L_\text{word}(w_j)$: & $-\log s_{\text{w}}(w_1)$ & -- & -- & $-\log  c_{\text{w}}(w_2)$\\
 \bottomrule
\end{tabular}
\vspace{-0.5em}
\end{table}
Although word-level deletions are not modeled, the subword/WP deletion problem is implicitly addressed by the loss. Table~\ref{Table: Edit} shows an example where the first word in the hypothesis contains a WP deletion. WP edit distance completely ignores any WP deletions because deletions are not part of the hypothesis. On the other hand, because the word-level CEM is trained with the word-level edit distance serving as the ground truth, hypothesized words that contain WP deletions are guaranteed to be labeled as word-level substitutions. This resolves an overconfidence problem that is common in token-level CEMs trained to model WP edit distance, especially for a smaller set of tokens such as graphemes.

\subsection{Deletion Length Prediction}
\label{Sec: Deletion}
To estimate the WER, the model needs to estimate the number of deletions as a secondary task. Here, we define a sequence of deletion length random variables for $1 \leq j \leq L+1$ as:
\begin{align*}
    E_j =
    \begin{cases}
      \text{\#(del) before the first word}, & \text{if } j = 1 \\
      \text{\#(del) after the last word}, & \text{if } j = L+1 \\
      \text{\#(del) between $j-1$ and $j$-th word}, & \text{otherwise}.
    \end{cases}    
\end{align*}
The number of deletions can be any non-negative integer. Existing works such as~\cite{ragni2018confidence} converts the problem into binary prediction by clipping $E_j$ at $1$.
In contrast, we predict $E_j$ directly by modeling $P(E_j|f(w_j);\theta)$ to follow the Poisson distribution, which is well-suited for count data. We use a MLP to parameterize it and predict the mean of $E_j$ as below:
\begin{align}
   \vspace{-1em}
   \mathbbm{E}(E_j|f(w_j);\theta) &= \exp(r_w(w_j)) \\
   r_w(w_j) &= \mathrm{MLP}\left( f(w_j); \theta \right) \label{Eq: Deletion MLP}
\end{align}
Let $e_j$ be the ground truth realization of $E_j$. With maximum likelihood estimation for Poisson regression, the training loss for deletions is given by:
\begin{align}
    \mathcal{L}_{\text{deletion}} &= -\sum_{j=1}^{L+1} \left[e_j r_w(w_j) - \exp{(r_w(w_j))}\right]. \label{Eq: Deletion loss}
\end{align}
During inference, the WER estimate can be computed as
\begin{align*}
    \uttwer &= \frac{\widehat{D}+\widehat{I}+\widehat{S}}{L+\widehat{D}-\widehat{I}},
\end{align*}
where $\widehat{D}=\sum_{j=1}^{L+1} \exp(r_{\text{w}}(w_j))$, $\widehat{I}=\sum_{j=1}^L i_{\text{w}}(w_j)$, and $\widehat{S}=\sum_{j=1}^L s_{\text{w}}(w_j)$.

\subsection{Utterance-Level Confidence Estimation}
A coarse way of including deletion information is to define a tertiary utterance-level prediction task, with the ground truth
\vspace{-0.5em}
\begin{align*}
    e_{\text{utt}} =
    \begin{cases}
      1, & \text{if utterance WER} = 0 \\
      0, & \text{otherwise}.
    \end{cases}
\end{align*}
The presence of any deletions causes the value of $e$ to be $0$, and this signal is backpropagated to the internal features in the Transformer feature extractor to help the word-level confidence task. This objective was proposed in~\cite{kumar2020utterance}, albeit in a single-task setup that ignores word confidences.
Intuitively, an utterance feature is a summary of the sequence of WP features. Thus, to extract the utterance features and make the prediction, we use the hierarchical attention proposed in~\cite{yang2016hierarchical}:
\begin{align}
    u_i &= \tanh{(W_1 f(y_i) + b)} \label{Eq: Hierarhical 1} \\
    \alpha_i &= \frac{\exp{(w_2^T u_i)}}{\sum_{m=1}^M \exp{(w_2^T u_m)}} \label{Eq: Hierarhical 2} \\
    \uttprob &= \sigma\left(\mathrm{MLP}\left(\sum_{i=1}^M \alpha_i f(y_i)\right)\right). \label{Eq: Hierarhical 3}
\end{align}
The parameters $W_1$, $b$, and $w_2$ that generate $\uttprob$ can be trained with the binary cross-entropy loss:
\begin{align}
\mathcal{L}_{\text{utt}} = -[e_{\text{utt}} \log \uttprob + (1-e_{\text{utt}})\log(1-\uttprob)]. \label{Eq: Utt loss}
\end{align}
$\uttprob$ provides an estimate of the probability that the entire utterance is recognized with zero WER, which is useful for ranking utterances in the order of their confidences. 

\subsection{Multi-task Objective}
The model can be trained by combining~\eqref{Eq: Word loss}, \eqref{Eq: Deletion loss}, and~\eqref{Eq: Utt loss}:
\begin{align*}
    \L_\text{total} = \frac{1}{L} \L_\text{word} + \frac{\lambda_\text{deletion}}{L+1} \L_\text{deletion} + \lambda_\text{utt} \L_\text{utt} .
\end{align*}
In this paper, $\lambda_\text{deletion} = 0.5$, $\lambda_\text{utt} = 1$,  when the respective losses are active, and $\L_\text{total}$ is further normalized over the mini-batch.

\section{Experimental Setup}
\subsection{Model Architecture}
The ASR's architecture is a RNN-T that follows~\cite{sainath2020streaming}. It has 8 LSTM layers in the encoder and 2 LSTM layers in the prediction network. Each LSTM layer is unidirectional, with 2,048 units and a projection layer with 640 units. A time-reduction layer is added between 2nd and 3rd encoder layers, which halves the sequence length by concatenating every two input frames. The RNN-T is frozen during CEM training. 

\begin{table}[ht]
\caption{Training objectives for five proposed models.}
\label{Table: Objective}
    \centering
    \begin{tabular}{ccccc}
        \toprule
         \multirow{2}{*}{Model} & \multirow{2}{*}{Model Size} & \multicolumn{3}{c}{Loss} \\
          \cmidrule{3-5}
           &  & $\L_\text{word}$ & $\L_\text{deletion}$ & $\L_\text{utt}$  \\
        \midrule
         \W & 16.6M & \cmark & \xmark & \xmark \\
         \U & 17.0M & \xmark & \xmark & \cmark \\
         \WD & 16.9M & \cmark & \cmark & \xmark \\
         \WU & 17.2M & \cmark & \xmark & \cmark \\
         \WUD & 17.5M & \cmark & \cmark & \cmark \\
        \bottomrule
    \end{tabular}
    \vspace{-2em}
\end{table}

We report the performance on the five combinations of confidence training tasks shown in Table~\ref{Table: Objective}. 
The feature extractor in~\eqref{Eq: Feature} consists of two Transformer decoder blocks with input embedding and internal dimensions of 640. For $\L_\text{deletion}$, the MLP in \eqref{Eq: Deletion MLP} has three layers, with hidden and output dimensions of 320, 160, and 1. For $\L_\text{utt}$, we use the hierarchical attention mechanism in \eqref{Eq: Hierarhical 1} and \eqref{Eq: Hierarhical 2} to generate the utterance-level feature. The MLP in \eqref{Eq: Hierarhical 3} has two layers, with hidden and output dimensions of 320 and 1, respectively.
All training is performed in TensorFlow with the Lingvo~\cite{shen2019lingvo} toolkit, using four hypotheses from the RNN-T. The optimizer is Adam~\cite{KingmaB14} with learning rate 0.001, and the global batch size is 4,096 across $8 \times 8$ TPU.

\subsection{Dataset}
The models are trained on the multi-domain training set used in~\cite{sainath2020streaming}, which spans domains of search, farfield, telephony and YouTube. All datasets are anonymized and hand-transcribed. The transcription for YouTube utterances is done in a semi-supervised fashion~\cite{liao2013large}. Multi-condition training (MTR)~\cite{kim2017generation} and random data downsampling to 8kHz~\cite{li2012improving} are applied to increase training data diversity.

The main test set includes $\sim$14K \textit{Voice Search} (VS) utterances extracted from Google traffic, which is anonymized and hand-transcribed. To test the generalization of the CEM,
we create a list of proper nouns identified by an in-house named entity tagger that are common in the LM training data for conventional ASR but rare in the audio-text paired training data for E2E ASR models. We select 10K sentences from the LM test data for the Maps domain, such that each sentence contains at least one of these proper nouns, then synthesize the TTS audio (as in~\cite{gonzalvo2016recent}) for them to create the \textit{Long-tail Maps} test set.
\color{black}

\subsection{Evaluation Metrics}
\begin{table*}[t]
    \caption{Confidence metrics on Voice Search and a long-tail Maps dataset for all proposed CEMs.}
    \label{Table: Metrics}
    \centering
    \begin{tabular}{ccccccccccccccccc}
        \toprule
          & \multicolumn{7}{c}{Voice Search} && \multicolumn{7}{c}{Long-tail Maps} \\
          & \multicolumn{3}{c}{Word-level} & & \multicolumn{3}{c}{Utterance-level} && \multicolumn{3}{c}{Word-level} & & \multicolumn{3}{c}{Utterance-level} \\
          \cmidrule{2-4}\cmidrule{6-8}\cmidrule{10-12}\cmidrule{14-16}
          & \multirow{2}{*}{NCE} & AUC & AUC & & AUC & AUC & (1-WER) && \multirow{2}{*}{NCE} & AUC & AUC & & AUC & AUC & (1-WER) \\
         Model & & ROC & PR & & ROC & PR & RMSE && & ROC & PR & & ROC & PR & RMSE \\
        \midrule
         \W & 0.348 & 0.927 & 0.451 && 0.765 & 0.428 & 0.226 && 0.285 & 0.875 & 0.494 && 0.704 & 0.549 & 0.301 \\
         \U & -- & --	& -- && 0.788 & 0.515 & -- && -- & -- & -- && 0.721 & 0.593 & -- \\ \hline
         \WD & 0.350 & 0.928 & 0.459 && 0.767 & 0.440 & 0.216 && 0.291 & 0.875 & 0.504 && 0.704 & 0.558 & 0.296 \\
         \WU & \textbf{0.378} & \textbf{0.931} & \textbf{0.489} && \textbf{0.810} & 0.549 & 0.223 && \textbf{0.304} & \textbf{0.883} & \textbf{0.524} && \textbf{0.755} & \textbf{0.636} & 0.296 \\
         \WUD & 0.365 & 0.929 & 0.476 && \textbf{0.810} & \textbf{0.552} & \textbf{0.213} && \textbf{0.304} & 0.882 & 0.516 && 0.746 & 0.634 & \textbf{0.292} \\
        \bottomrule
    \end{tabular}
    \vspace{-.1in}
\end{table*}

All evaluation metrics are reported on the top hypothesis. For word confidence, we adopt the same metrics in~\cite{qiu2021learning}: normalized cross-entropy (NCE)~\cite{Siu1997ImprovedEE}, area under the receiver operating characteristic curve (AUC-ROC) and the \textit{negative} class precision recall curve (AUC-PR). Higher is better for these metrics.

To measure the ranking quality of the utterance confidence, we use the utterance-level AUC-ROC and AUC-PR (higher is better), where the positive / negative classes are the sets of utterances with zero / non-zero WER, respectively. To measure the accuracy of estimating WERs, we use the root mean squared error (RMSE, lower is better) between the utterance confidence and the ground truth $(1-\text{WER})$.

\begin{table}[ht]
\caption{Utterance confidence scores' priorities by metrics.}
\label{Table: Priority}
    \centering
    \begin{tabular}{cccc}
        \toprule
          Model & Available Scores & AUC-ROC \& PR & RMSE  \\
        \midrule
         \W & $\uttwcr$ & $\uttwcr$ & $\uttwcr$ \\
         \U & $\uttprob$ & $\uttprob$ & N/A \\
         \WD & $\uttwcr$, $\uttwer$ & $1-\uttwer$ & $1-\uttwer$ \\
         \WU & $\uttwcr$, $\uttprob$ & $\uttprob$ & $\uttwcr$ \\
         \WUD & $\uttwcr$, $\uttprob$, $\uttwer$ & $\uttprob$ & $1-\uttwer$ \\
        \bottomrule
    \end{tabular}
    \vspace{-1em}
\end{table}

Table~\ref{Table: Priority} shows the availability of utterance-level confidence scores, and which one is preferred when multiple are available. Recall that $\uttprob$ models the probability that the utterance is recognized with zero WER. Thus, we prioritize using it for the utterance ranking tasks (AUC-ROC, AUC-PR). However, $\uttprob$ does not attempt to estimate the WER, and we de-prioritize it for the WER accuracy task (RMSE).

\section{Results}
\subsection{Joint Training Confidence Metrics}

This section discusses the performance of all models on the confidence metrics. In Table~\ref{Table: Metrics}, comparing Models \WU{}, \WD{}, and \WUD{} vs Model \W{} shows that any joint training (especially with the utterance-level loss) helps improve word-level metrics. Comparing Models \WU{} and \WUD{} vs Model \U{} shows that joint training also helps improve the performance of $\uttprob$ on utterance-level AUC-ROC. Comparing Models \WD{} and \WUD{} vs Model \W{} shows that modeling deletion and insertion rates helps to achieve the most accurate WER estimate in terms of RMSE. Since all five models have similar number of parameters (see Table~\ref{Table: Objective}), it is clear that the three tasks do not significantly increase the required model capacity, and their training signals complement each other to improve overall confidence accuracy.

\subsection{Qualitative Example}
To visualize the benefit of adding $\L_\text{deletion}$ and $\L_\text{utt}$, Table~\ref{Table: Example} shows an example utterance that contains a deletion. Model \WUD{}, through learning the utterance correctness and deletion lengths at every position, is able to use that knowledge to aid word confidence predictions. That is seen from the lower confidence score at the word ``absolut'', which follows a deletion.

\begin{table}[!ht]
\vspace{-.1in}
\caption{The CEM output $c_w(w_j)$ on an example TTS utterance, for Models \W{} and \WUD{}.}
\label{Table: Example}
    \centering
    \begin{tabular}{l@{\hspace{1.3\tabcolsep}}cccccc}
        \toprule
        Hyp: & & absolut & green & philadelphia & pa \\
        Ref:  & upsal & at & greene & philadelphia & pa \\ 
        Word & \multirow{2}{*}{\textit{del}} & \multirow{2}{*}{\textit{sub}} & \multirow{2}{*}{\textit{sub}} & \multirow{2}{*}{\textit{cor}} & \multirow{2}{*}{\textit{cor}} \\
        edit: \\
        \midrule
        \W{}:    & -- & 0.62 & 0.95 & 0.99 & 0.99 \\
        \WUD{}: & -- & 0.24 & 0.92 & 0.99 & 0.88 \\
        \bottomrule
    \end{tabular}
    \vspace{-1em}
\end{table}

\subsection{Rescoring Using Utterance Confidence}

\begin{figure}[ht]
    \centering
    \begin{subfigure}{.48\linewidth}
      \centering
      \includegraphics[width=\textwidth]{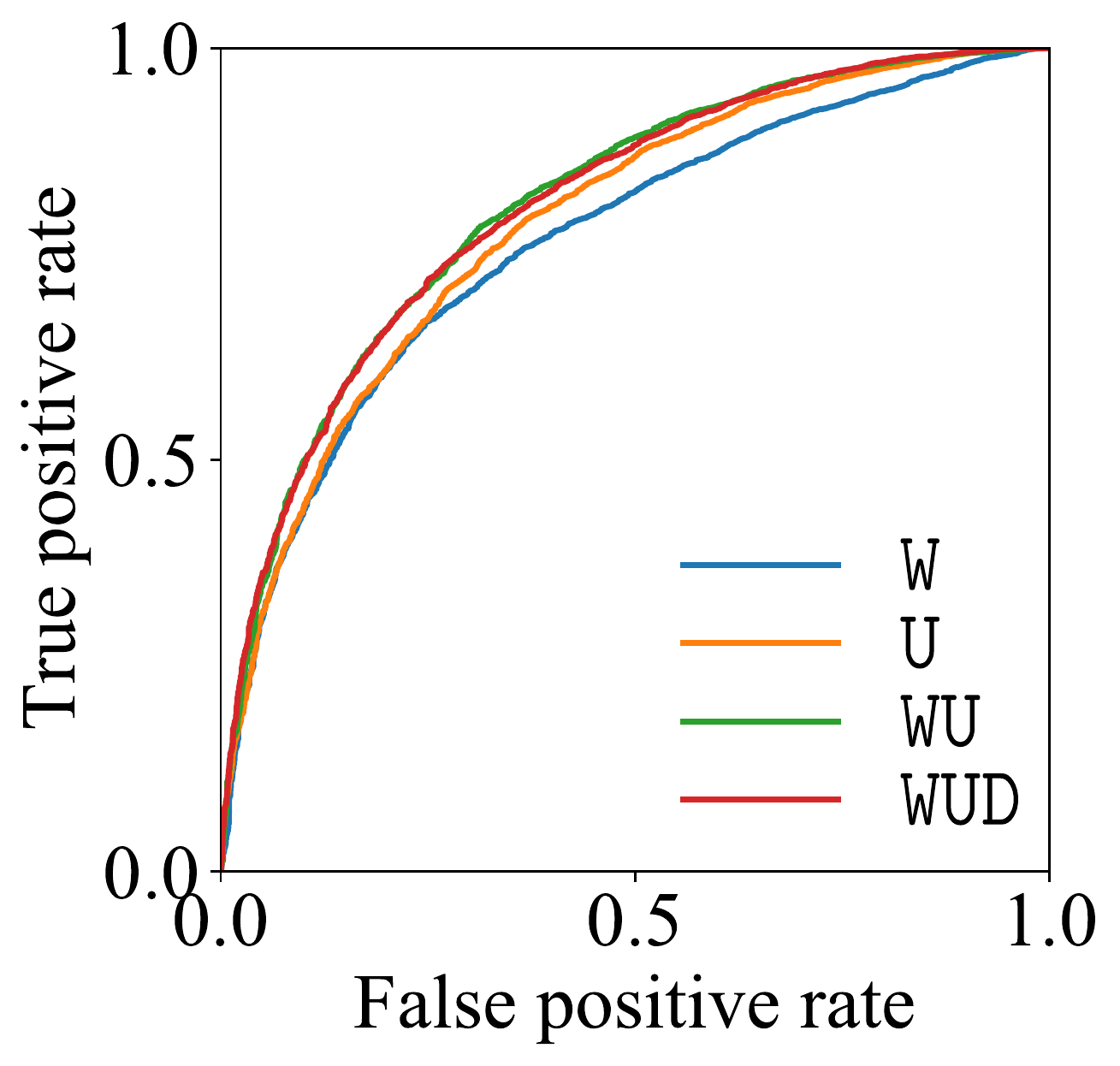}
      \caption{ROC cuves.}
      \vspace{-0.5em}
      \label{fig:roc}
    \end{subfigure}%
    \begin{subfigure}{.48\linewidth}
      \centering
      \includegraphics[width=\textwidth]{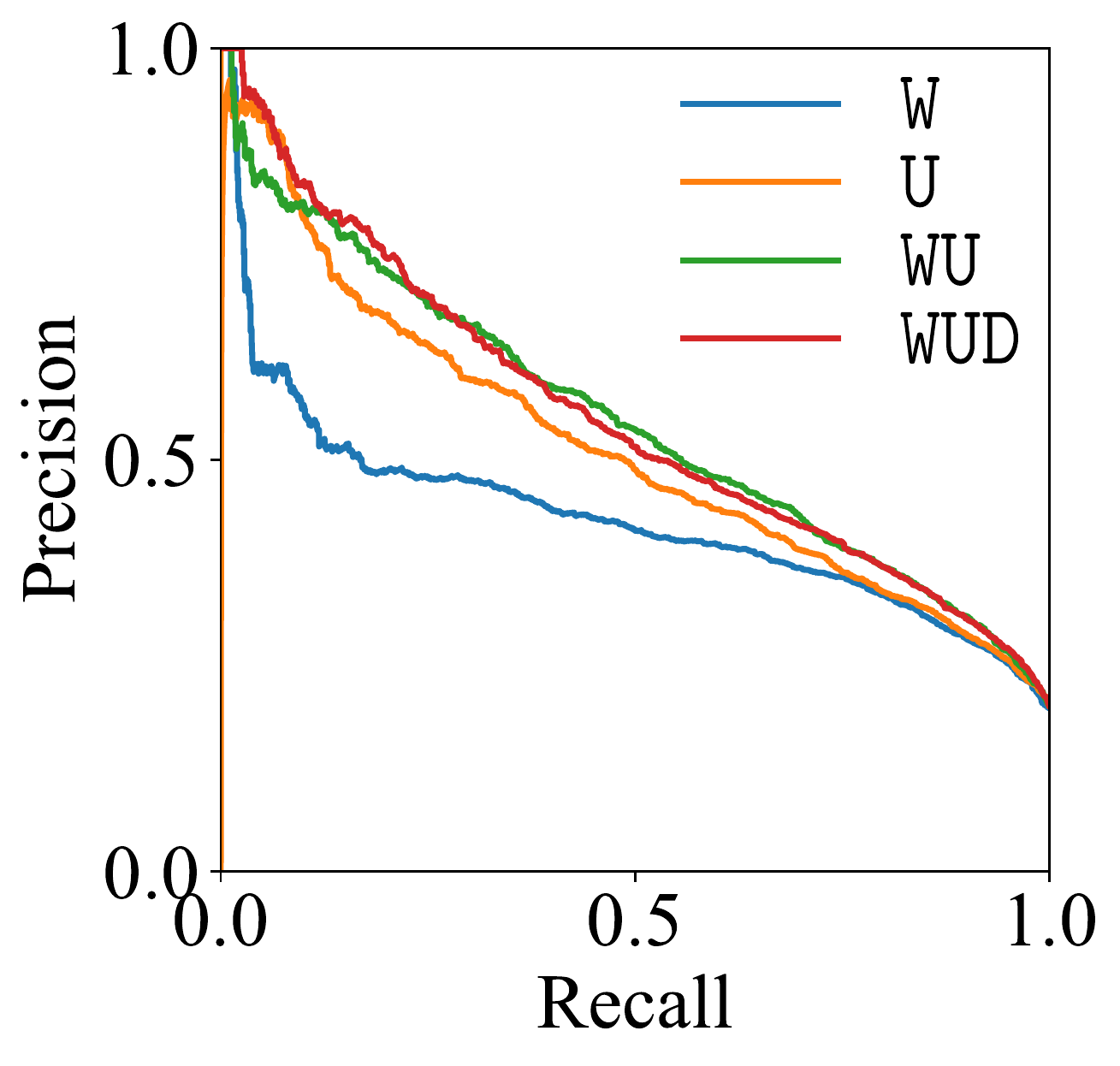}
      \caption{PR curves.}
      \vspace{-0.5em}
      \label{fig:pr}
    \end{subfigure}
    \caption{Utterance-level ROC and PR curves on Voice Search for Models \W{}, \U{}, \WU{}, \WUD{}.}
    \label{Fig: ROC}
    \vspace{-1em}
\end{figure}

Figure~\ref{Fig: ROC} shows the improved utterance-level ROC and PR curves for the multi-task Model \WU{} over single-task Models \W{} and \U{}. This improvement in utterance-level confidence scores can serve as an alternative signal to determine which RNN-T beam search hypothesis has the lowest WER among alternative hypotheses in the beam. For every utterance, we use confidence to rescore the top four RNN-T hypotheses. In Table~\ref{Table: Rescoring}, we report the average WER on the top hypothesis after rescoring. On Voice Search, both the $\uttwcr$ and $\uttprob$ outputs of Model \WU{} reduce the Voice Search WER from 6.4 to 6.1, which shows the advantage of joint training. On the long-tail Maps set, $\uttwcr$ achieves the lowest WER of 13.6, which suggests that word-level confidence shows better generalization to previously unseen data.

\begin{table}[ht]
\caption{Rescoring WERs (\%).}
\vspace{-.5em}
\label{Table: Rescoring}
    \centering
    \begin{tabular}{cc|cc}
        \toprule
          Model & Rank & Voice Search & Long-tail Maps \\
        \midrule
        RNN-T & Beam search & 6.4 & 14.0 \\
        \midrule 
        \W   & $\uttwcr$ & 6.2 & 13.8 \\
        \U   & $\uttprob$ & 6.3 & 14.4 \\
        \WU  & $\uttwcr$ & \textbf{6.1} & \textbf{13.6} \\
        \WU  & $\uttprob$ & \textbf{6.1} & 14.1 \\
        \WUD & $\uttwcr$ & 6.2 & \textbf{13.6} \\
        \WUD & $\uttprob$ & 6.2 & 14.1 \\
        \bottomrule
    \end{tabular}
    \vspace{-1em}
\end{table}

\section{Conclusions}
In this paper, we propose to extend the word-level CEM to jointly learn the length of deletions between every pair of words and the probability that the entire utterance is recognized correctly. Experimental results show that the proposed multi-task learning setup improves word and utterance-level confidence metrics without incurring significant model size increase. The utterance confidence can also be used to rescore first-pass RNN-T hypotheses and reduce WER by 3-5\% relative.

\newpage
\bibliographystyle{IEEEtran}

\bibliography{main}

\end{document}